\documentstyle{report}

\begin{document}

\bf{POSSIBLE IMPLICATIONS OF THE QUANTUM THEORY OF GRAVITY}

\smallskip

An Introduction to the Meduso-Anthropic Principle

\smallskip

by Louis Crane, math department, KSU

\smallskip

In his excellent book ``Life and Light'' (still in press as of this
writing) Lee Smolin has proposed that if the quantum theory of
gravity has two special features the universe would fine tune itself.

Specifically, he proposes [1] that the quantum theory softens the
singularity which forms in a Kerr-Newmann black hole, for example,
if we perturb it, so that the new universe on the other side of it will
really form. He then proposes [2] that the constants of nature might
fluctuate in such a process, so that the new universe would have
slightly different physics in it. In the absence of a quantum theory
of gravity, of course, both of these suggestions are speculative.

Nevertheless, it is interesting to see what would result from
such a process. Universes with peculiar fine tuned values of the
coupling constants would have many more ``daughter'' universes,
so that fine tuning, instead of an improbable accident, would
be highly probable.

Professor Smolin proposes this as an explanation of the fine tuning
evident in our universe, which leads to the possibility of life.
In this view, life and intelligence are biproducts of the very special
physics necessary to provide surface chemistry and radiating stars
in order to produce many generations of black holes.

Philosophically, then, life and even intelligent life are accidents.

I do not intend to repeat Professor Smolin's argument in this note.
Rather, I wish to propose a modification of his reasoning which
seems practically necessary in the framework  of his conjectures,
but which changes their philosophical implications immensely.

I have called this modified version the meduso-anthropic principle,
after a stage in the life cycle of the jellyfish. The reason for this
metaphor will be explained in time.

In addition, I want to point out in this note that within the
approach to quantizing gravity which I proposed in my paper
``Topological Quantum Field Theory as The Key to Quantum Gravity''
the second conjecture which Professor Smolin makes, ie that coupling
constants might fluctuate when the topology of spacetime changes,
becomes much more plausible. This is because I propose a quantum
theory in which the coupling constant is part of the state of the
universe. If this approach were extended to gravity coupled to matter,
the coupling constants of the matter fields would have a similar role.
One could try to compute topology changing amplitudes
from them via the state sum methods in my paper. I do not yet see how
to do this, but there are some suggestive possibilities in the underlying
algebraic picture.

The question whether the real quantum theory of gravity has the right
features to make this picture work will no doubt remain open for
a considerable time. Nevertheless, let us explore the implications.

\bigskip

\bf{MAN IN THE LOOP}

\bigskip

The conjecture which I believe modifies Professor Smolin's conclusions
is the following:

\smallskip

SUCCESSFUL INDUSTRIAL CIVIOLIZATIONS WILL EVENTUALLY CREATE BLACK
HOLES.

\smallskip

The synthesis of this with Smolin's two conjectures is what I call
the meduso-anthropic principle. Before exploring the implications,
let us consider the plausibility of this conjecture.

\smallskip

SUBCONJECTURE 1: SUCCESSFUL INDUSTRIAL CIVILIZATIONS WILL EVENTUALLY
WANT TO MAKE BLACK HOLES

\smallskip

and

\smallskip

SUBCONJECTURE 2: SUCCESSFUL INDUSTRIAL CIVILIZATIONS WILL EVENTUALLY
BE ABLE TO PRODUCE BLACK HOLES.

\smallskip

It is fairly clear, at least, that the conjecture follows from the
two subconjectures. (This paper is not on the mathematical level of
rigor).

Let us first consider subconjecture 1. There are two reasons to want
to make black holes. One might want to make a few for scientific
purposes. Indeed, barring major surprizes in physics they are the
ultimate high energy phenomenon. If it came within the reach of
a technological civilization to build them, certainly the scientists
in it would want to do so.

The second motivation for creating black holes is much more
compelling. The hydrogen supply of the universe is slowly being
exhausted. At some point in the future, any civilization will face
the possibility of perishing from a lack of free energy, In principle,
black holes could provide a permanent solution to this problem, since
they can convert matter into energy via the Hawking radiation forever
and with perfect efficiency. (They are also the perfect waste disposal
for similar reasons). In order to make this practical, it would be
necessary to have very small and very hot black holes, and to be able
to ``feed'' and manage them very carefully. However difficult this
problem finally is, our descendants in a few hundred billion years
will have no alternative if they want to go on living.

Now let us consider the second subconjecture. The main difficulty in
creating a black hole is cramming a lot of mass-energy in a small
space.

Nature solves this problem by cramming a lot of nuclear matter into
the center of a large star. This is completely inadequate for
our purposes, since the resulting black holes are much too big,
and hence much too cold to be of use. Also, it is hard to imagine
a civilization doing such a thing.

Fortunately, there are two approaches which could produce much higher
densities, and hence much smaller holes.

In one approach, one simply creates a huge sphere of converging lasers
and fires them simultaneously at a central point. Since light is
composed of bosons, there is no Pauli exclusion principle to overcome,
and the bursts of photons could all occupy a very small space
simultaneously,
creating a black hole of a temperature corresponding to the frequency
of the light. (The term ``successful' in the conjecture has to be
taken on such a scale. Still, a hundred billion years is a long time).

The critical length in such an apparatus is the wavelength of the
light used. If our descendents can build nuclear lasers which lase in
hard gamma, then a spherical converging laser the size of a small
asteroid would suffice to produce very small hot black holes. Of
course,
gamma interferometry would be necessary  to keep it focussed. None of
this is beyond what could plausibly be done in a few centuries.

Another approach involves ordinary fermionic matter. One creates very
long thin
cylinders  and accelerates them towards one another at high
relativistic velocity. This gets around the density problem for two
reasons: first, the cylinders would be lorentz contracted, and second,
their rest masses would increase by a gamma factor. The combination of
these two effects would produce a very large effective compression if
ultrarelativistic velocities could be reached.

Both of these approaches pose subtle relativity questions, as well as
extremely obvious engineering ones. Nonetheless, I believe that they
demonstrate some degree of plausibility for my second subconjecture.

The notion of a successful civilization here is considerably beyond
contemporary standards. The picture I am imagining is of a
civilization which has reached a galactic center or some other stellar
cluster, and is able to rework the resources of an entire solar system
into a single huge machine. The energy output of the entire star could
be diverted to powering the machine. This is not so outlandish as it
first appears, given robotic factories.

\bigskip

\bf{IMPLICATIONS OF THE CONJECTURES}

If both Smolin's two conjectures and mine are true, then the fine
tuning of physical constants would not stop with physics which
produced stars and gas clouds which cool. Rather, the selection would
continue until physics evolved which resulted in successful
civilizations, with a very exacting definition of success. In the
limit as the number of generations of daughter universes increases,
this would be true even if we only make the weaker assumption that
successful civilizations make a few black holes as experiments. This
would increase the average number of daughter universes in universes
with successful civilizations, even if by a small fraction. Each such
universe would have more daughters, until, after many generations,
intelligences would be present in almost all universes. The effect
would be much more rapid if black holes as energy sources turn out to
be practical.

The philosophical implications of such a process are very deep.
Although it has been generally believed by people with a scientific
frame of mind that human life and history take place within the rule
of physical law, it has generally been assumed that the relationship
between the specific laws of physics and human events was complex and
accidental. This has, in fact placed science in conflict with the
otherwise dominant currents of Western (and by no means only Western) thought.

Indeed, it has been the belief of most philosophers, and a surprizing
number of important scientists, that humanity had some fundamental
role in the universe, and that mind was more than an accidental
attribute of organized matter.

If the combination of hypotheses described above is correct, a richer
connection between mind and matter appears in a surprizing way. Almost
all universes would produce successful intelligence, because their
detailed structure would be fine tuned by a long process in which
intelligences had reproduced universes over and over; a process with
the closest analogy with the passage of millions of generations which
has honed life forms to an almost unimaginable perfection.

If the combination of hypotheses which I am giving the name of
meduso-anthropic is correct, the relationship of civilization to
environment would entail a thousand improbable coincidences with
favorable
outcomes. Historical events would skirt innumerable disasters and
find an improbable path to success. The relationship of humanity to
the
universe would have  an organic quality.

It is now possible to explain the metaphor I have chosen in the title
meduso-anthropic. It refers to a stage in the development of the
animals in the phylum which includes the jellyfish and coral.
These animals have two phases of life, medusid and polyp. Medusids
produce polyps, which produce medusids. It is sometimes even difficult
to recognize that the two stages represent a single species.
Analogically, intelligences are the medusids, and black
holes/universes are the polyps of a single evolutionary process.

The idea of an organic fine tuning of the relationship of life to
nature seems improbable as long as one listens to the voice of
scientific common sense. The minute one examines any part of natural
history as our understanding of it is growing, experience begins to
drown out common sense. How did we really end up with a repeating
series of extinctions, frequent enough to drive on evolution, yet rare
enough to permit it? Why is there so much fossil fuel on the earth?
For that matter, why didnt we discover the atom bomb a few years
earlier, or fight world war 2 a few years later?

It is not hard to see that if these ideas are true, they will be the
victims of abuse to dwarf quantum healing and even quantum golf. That
is not sufficient reason to ignore them.

\bigskip

A SUMMATION

\bigskip

It is certainly impossible to claim that the quantum theory of gravity
has reached a stage where these ideas can be validated. On the other
hand,
it could reach such a stage in fairly short order, if we are lucky.

Nevertheless this much seems clear. The laws of Physics which only
operate at very high energies can nevertheless have profound
implications for what we see around us. They have the possibility of
changing our understanding of ourselves and our world in ways we
cannot yet imagine. The fact that machines at Planck scale energies
are not yet in the cards does not mean that quantum gravity is of no
concern to us.

For myself, I intend to return to my four dimensional state sums with
greatly increased ardor, in the hope that something like the
meduso-anthropic principle might emerge from them.

\end{document}